\documentclass[12pt]{article}
\usepackage{amsmath,amssymb,color}
\usepackage{cite,epsf}
\usepackage{pstricks}
\usepackage{color}

\usepackage{graphicx,array} 
\newcommand{\be}{\begin{equation}}
\newcommand{\ee}{\end{equation}}
\newcommand{\beq}{\begin{equation}}
\newcommand{\eeq}{\end{equation}}
\newcommand{\bea}{\begin{eqnarray}}
\newcommand{\eea}{\end{eqnarray}}

\newcommand{\ba}{\begin{eqnarray}}
\newcommand{\ea}{\end{eqnarray}}

\def\sin{\mbox{sin}}
\def\cos{\mbox{cos}}
\def\log{\mbox{log}}

\begin{document}

\begin{titlepage}
\vspace{10pt}
\hfill
{\large\bf HU-EP-15/38}
\vspace{20mm}
\begin{center}

{\Large\bf  Wilson loops at strong coupling \\[2mm]
for curved contours with cusps\\[2mm] 
}

\vspace{45pt}

{\large Harald Dorn 
{\footnote{dorn@physik.hu-berlin.de
 }}}
\\[15mm]
{\it\ Institut f\"ur Physik und IRIS Adlershof, 
Humboldt-Universit\"at zu Berlin,}\\
{\it Zum Gro{\ss}en Windkanal 6, D-12489 Berlin, Germany}\\[4mm]

\vspace{20pt}

\end{center}
\vspace{10pt}
\vspace{40pt}

\centerline{{\bf{Abstract}}}
\vspace*{5mm}
\noindent
We construct the minimal surface in $AdS$, relevant for the strong coupling behaviour
of local supersymmetric Wilson loops in ${\cal N}=4$ SYM for a closed contour formed
out of segments of two intersecting circles. Its regularised area is calculated
including all divergent parts and the finite renormalised term. Furthermore we prove, that for generic planar curved contours with cusps  the cusp anomalous dimensions are functions of the respective cusp angles alone. They do not depend on other local data of the cusps. 
\vspace*{4mm}
\noindent

\vspace*{5mm}
\noindent
   
\end{titlepage}
\newpage


\section{Introduction}
The renormalisation properties of Wilson loops in perturbative gauge theories
are well understood for a long time \cite{Polyakov:1980ca,Gervais:1979fv,Dotsenko:1979wb,Brandt:1981kf,Dorn:1986dt}. With smooth closed contours
there appears an exponential of a linear divergence proportional to the length of the contour. Besides removing this factor and the renormalisation of the coupling constant, no further renormalisation is needed to get finite results. This changes if the contour has cusps or self-intersections, where additional logarithmic divergences
show up. The corresponding cusp anomalous dimension has been calculated
for QCD in one \cite{Polyakov:1980ca}, two \cite{Korchemsky:1987wg} and three loop order \cite{Grozin:2014hna}.

In ${\cal N}=4$ SYM the situation improves. At first the theory is conformally invariant
and the coupling needs no renormalisation, and for local supersymmetric
loops the linear divergences cancel \cite{Drukker:1999zq}. The logarithmic cusp
divergences are still present and of great interest for a lot of applications, a recent calculation up to the four loop order in planar ${\cal N}=4$ SYM one finds in \cite{Henn:2013wfa}.
 
But even more important, with the $AdS$/CFT
correspondence one has a handle to the strong coupling behaviour.
At strong 't Hooft coupling $\lambda$, the locally supersymmetric Wilson loop for a 
closed contour ${\cal C}$ is in leading order given by  \cite{Maldacena:1998im,Rey:1998ik} 
\beq
W({\cal C})~=~\mbox{exp}\Big (-\frac{\sqrt{\lambda}}{2\pi}~A({\cal C})\Big )~.\label{W-A}
\eeq
$A({\cal C})$ is the area of a minimal surface extending into the bulk of $AdS_5$
and approaching its boundary along the contour ${\cal C} $. The construction uses
Poincar{\' e} coordinates
\beq
ds^2~=~\frac{1}{r^2}\big (dx^{\mu}dx_{\mu}~+~dr^2 \big )~,\label{poincare}
\eeq 
where the conformal boundary is at $r=0$. The UV divergences of the gauge theory 
reappear as divergences due to the blow up of the metric near $r=0$. The standard procedure to define a regularised area  $A_{\epsilon}$ is based on cutting off that part of the surface on which $r<\epsilon$. 

For the interpolation between weak and strong coupling, integrability techniques have been used
in \cite{Correa:2012hh} to derive a set of equations for the cusp anomalous dimension.

Solving the minimal surface condition (equation of motion, if seen as string surface)
near the boundary of $AdS$, one gets for space-like contours \cite{Graham:1999pm,Polyakov:2000ti}
\beq
x^{\mu}(\sigma,r)~=~x^{\mu}(\sigma,0)~+~\frac{1}{2} \frac{d^2}{d\sigma ^2}x^{\mu}(\sigma,0)~r^2~+~{\cal O}(r^3)~,\label{surface-asympt}
\eeq
where $\sigma$ together with the fifth $AdS$ coordinate $r$ parameterises the surface and is the length parameter on the boundary curve $x^{\mu}(\sigma,0)$. The absence of a term linear in $r$ 
in eq.\eqref{surface-asympt} ensures that for smooth boundary curves with their bounded curvature
one gets ($l$ length of the Wilson loop contour)
\beq
A_{\epsilon}~=~\frac{l}{\epsilon}~+~A_{\mbox{\tiny ren}}~+~{\cal O}(\epsilon)~.
\eeq 
The invariance of $A_{\mbox{\tiny ren}}$ under conformal rescalings of the metric
on the boundary of $AdS$ has been proven in \cite{Graham:1999pm}. The equivalent
issue of keeping the metric fixed, but performing an active conformal mapping of 
the boundary points has been discussed on the level of infinitesimal transformations
in \cite{Muller:2013rta}. A direct proof for finite transformations can be given
by using \eqref{surface-asympt} and applying Stokes theorem on that part of the original surface
bounded by the line $r=\epsilon$ and the preimage of the line which is at the same value of $r$ on the 
mapped surface.

In the presence of cusps the expansion in \eqref{surface-asympt} with uniformly bounded coefficients breaks down. This is the reason for the appearance
of logarithmic divergences in $A_{\epsilon}$ for $\epsilon\rightarrow 0$. 
For a generic smooth contour with $n$ cusp one commonly expects  \footnote{To match the above
mentioned absence of linear divergences in the small coupling expansion one has
to understand the area $A$ in the regularised version of \eqref{W-A} already after
subtracting the $l/\epsilon$ term. This can be understood as the effect of a certain Legendre transformation \cite{Drukker:1999zq}.}
\beq
A_{\epsilon}~=~\frac{l}{\epsilon}~+~\sum _{i=1}^n~\Gamma_{\mbox{\tiny cusp}}(\theta_i)~\mbox{log}~\epsilon~+~A_{\mbox{\tiny ren}}+~{\cal O}(\epsilon)~.\label{div-structure}
\eeq

Now we are ready to pose the question treated in this paper. All calculations of
the coefficient in the logarithmic divergence have been performed for a cusp
with straight legs \cite{Drukker:1999zq,Kruczenski:2002fb}, and it is commonly believed that the above  structure, with the same $\Gamma_{\mbox{\tiny cusp}}$ depending only on 
the cusp angles $\theta _i$, is true also for cusps with curved legs. We want to prove that this belief is justified indeed.
Of course, it is clear that $\Gamma_{\mbox{\tiny cusp}}$ can depend on dimensionless local data only. But for generic cusps
there are  available, besides the angles,  e.g. the quotients of the right and left limits of the curvature.

In the case of the field theoretic small coupling expansion the use of the simplification
achieved by working with straight legs is easier to justify.
In comparison to the smooth case, these divergences are due to the change generated by the cusp into the projection of four-dimensional distances $(x-y)^2$ onto the one-dimensional contour parameter space.

The paper is organised as follows. In section 2 we calculate $A_{\epsilon}$ for contours formed out of segments of two intersecting
circles with arbitrary radii, including all divergent terms as well as $A_{\mbox{\tiny ren}}$. The result
for this example will fit into the structure \eqref{div-structure}. 

Section 3 is devoted to generic contours in an Euclidean plane. A  suitable surface parameterisation for generic cusps with curved legs is developed
as some kind of perturbation of that used in the case of straight legs in \cite{Drukker:1999zq}. Then on its
basis a general proof will be given.

Several technical details are collected in six appendices.
\section{Contour with two cusps, formed by segments of two circles}
Two straight half-lines, starting on the $x_1$-axis at $x_1=q>0$ with angles $\gamma _1< \gamma_2$, form a cusp with angle \footnote{We follow the convention used in 
\cite{Drukker:1999zq}, i.e. smooth case corresponds to $\theta=\pi$.}
\beq
\theta =\gamma_2-\gamma_1.
\eeq 
The corresponding minimal surface is given by \cite{Drukker:1999zq}
\beq
x_1~=~q~+~\rho  ~\mbox{cos}(\varphi +\gamma_1)~,~~~~~~~x_2~=~\rho ~\mbox{sin}(\varphi +\gamma_1)~,~~~~~~~
r~=~\frac{\rho}{f(\varphi)}~,\label{orignal-surface}
\eeq
with $0\leq\rho<\infty~,~~~0\leq\varphi\leq\theta$. The essential properties of this solution are
summarised in appendix B,
in particular $f(\varphi )$ is defined implicitely by eqs. \eqref{E-f},\eqref{phi-f}.

The map
\beq
x_{\mu}~\mapsto ~y_{\mu}~=~\frac{x_{\mu}}{x^2+r^2}~,~~~~~r~\mapsto ~z~=~\frac{r}{x^2+r^2}~\label{ads-trafo}
\eeq
is an isometry inside $AdS$ and acts as a conformal transformation on the boundary (inversion at the unit circle). It maps the two half-lines discussed above to segments of two circles forming a closed contour with two cusps, both with angle $\theta$
\footnote{A similar construction in the context of BPS Wilson loops composed  of two longitudes on a $S^2$ has been used in \cite{Drukker:2007qr}. }
. For some details see appendix A.

Due to the bulk isometry property of \eqref{ads-trafo}, the minimal surface related to this closed contour 
is then given by
\bea
y_1&=&\frac{\rho~\mbox{cos}(\varphi +\gamma _1)~+~q}{\rho ^2 +q^2+2q\rho~\mbox{cos}(\varphi+\gamma _1)+(\rho /
f(\varphi))^2}~,\nonumber\\[2mm]
y_2&=&\frac{\rho~\mbox{sin}(\varphi +\gamma _1)}{\rho ^2 +q^2+2q\rho~\mbox{cos}(\varphi+\gamma _1)+(\rho/ f(\varphi))^2}~,\nonumber\\[2mm]
z&=&\frac{\rho/f}{\rho ^2 +q^2+2q\rho~\mbox{cos}(\varphi+\gamma _1)+(\rho/ f(\varphi))^2}~.\label{mapped-surface}
\eea

For all surfaces we follow the standard definition of the regularised area, i.e. cutting off that part of the surface
whose Poincar{\' e} $r$-coordinate (see \eqref{poincare}) is smaller than $\epsilon $. Since we denoted this coordinate
for our surface generated by the map \eqref{ads-trafo} with the letter $z$, we have to calculate
\beq
A_{\epsilon}~=~\int _{z>\epsilon}\sqrt{h}~d\rho d\varphi\label{reg-area}~,
\eeq
where $h$ is the determinant of the induced metric on the surface \eqref{mapped-surface}. There is an alternative way
to get the same quantity: calculate the area of the preimage of the cut surface on the original \eqref{orignal-surface}.
Then for $h$ one has to take the simple form 
as in \eqref{ind-metric-det} of appendix B and the integration region
for $\rho$ at fixed allowed $\varphi$ is given by
\beq
\rho _-<\rho<\rho_+~,\label{rrr}
\eeq
with
\beq
\rho_{\pm}=\frac{1-2\epsilon qf(\varphi)\mbox{cos}(\varphi+\gamma_1)\pm \sqrt{1-4\epsilon qf\mbox{cos}(\varphi+\gamma_1)-4\epsilon ^2q^2f^2\mbox{sin}^2(\varphi+\gamma_1)-4\epsilon ^2q^2}}{2\epsilon (f+1/f)}\nonumber
\eeq
denoting the two solutions of the equation $z=\epsilon $.
The square root in the above formula has to be real. This gives  implicit bounds for $\varphi $ via
\beq
f_0~<~f(\varphi)~<~f_{\epsilon}~.
\eeq

To characterise $f_{\epsilon}$, one has to handle the fact that the relation between  $\varphi$ and $f $ is not one to one. Instead we
have  $f(\varphi)=f(\theta -\varphi)$, and in $\varphi\in (0,\theta/2)$ the function $\varphi(f)$ is given by formula \eqref{phi-f} of appendix B. Therefore, we split the regularised area $A_{\epsilon}$ into two pieces, one originating from $\varphi\in (0,\theta/2)$
and the other from $\varphi\in (\theta/2,\theta)$
\beq
A_{\epsilon}~=~A_{\epsilon}^{(1)}~+~A_{\epsilon}^{(2)}~. \label{AA1A2}
\eeq
Having in mind  $\varphi +\gamma_1=\gamma_2-(\theta-\varphi)$, we define $f_{\epsilon}^{(j)}$ ($j=1,2$) by
\beq
f_{\epsilon}^{(j)}~=~\frac{\sqrt{1-4\epsilon^2q^2\mbox{sin}^2\big (\gamma_j\pm\varphi(f_{\epsilon}^{(j)})\big)~}-\mbox{cos}\big (\gamma_j\pm\varphi(f_{\epsilon}^{(j)})\big)}{2\epsilon~ q~\mbox{sin}^2\big (\gamma_j\pm \varphi(f_{\epsilon}^{(j)})\big )}~.\label{fepsj}
\eeq
Estimating this implicit definition for $\epsilon\rightarrow 0$, which is correlated with $f_{\epsilon}^{(j)}\rightarrow\infty$ and via \eqref{phi-f-as} with $\varphi(f_{\epsilon}^{(j)})\rightarrow 0$ this simplifies to
\beq
f_{\epsilon}^{(j)}~=~\frac{T_j}{\epsilon}~-~q~+~{\cal O}(\epsilon)~,~~~~~\mbox{with}~~~~~T_j~=~\frac{1-\mbox{cos}\gamma_j}{2q ~\mbox{sin}^2\gamma_j}~.\label{Tj}
\eeq
After substituting the integration variables $\varphi$ or $\theta -\varphi$, respectively, via \eqref{phi-f} of appendix B by $f$, we
get for $j=1,2$
\bea
A_{\epsilon}^{(j)}&=&\int _{f_0}^{f_{\epsilon}^{(j)}} df\int _{\rho_-(f)}^{\rho _+(f)}\frac{d\rho}{\rho}\sqrt{\frac{f^4+f^2}{f^4+f^2-E^2}}\nonumber \\[2mm]
&=&\int _{f_0}^{f_{\epsilon}^{(j)}} df~\sqrt{\frac{f^4+f^2}{f^4+f^2-E^2}}~\Big (2~\mbox{log}\frac{\rho_+^{(j)}(f)}{q}~+~\mbox{log}\frac{1+f^2}{f^2}\Big )~,\\[2mm]\nonumber
\eea
where we used that \eqref{rrr} implies $\rho_+^{(j)}\rho_-^{(j)}=\frac{q^2f^2}{1+f^2}$. Here $\rho_{\pm}^{(1)}$ is given by  the r.h.s of  \eqref{rrr}
and $\rho_{\pm}^{(2)}$ after the replacement of $\gamma_1+\varphi(f)$ by $\gamma_2-\varphi(f)$.

Let us now separate $A_{\epsilon}^{(j)}$ into two additive pieces, where only the first one contains the function $\varphi(f)$
\bea
A_{\epsilon}^{(j)}&=&A_{\epsilon}^{(j,1)}~+~A_{\epsilon}^{(j,2)}~,\label{A1+A2}\\[2mm]
A_{\epsilon}^{(j,1)}&=&2~\int _{f_0}^{f_{\epsilon}^{(j)}} df~\sqrt{\frac{f^4+f^2}{f^4+f^2-E^2}}~\mbox{log}N_{\epsilon}^{(j)}(q,f)~,\label{defA1}\\
A_{\epsilon}^{(j,2)}&=&-~\int _{f_0}^{f_{\epsilon}^{(j)}} df~\sqrt{\frac{f^4+f^2}{f^4+f^2-E^2}}~\big (2\mbox{log}(2q\epsilon)~+~\mbox{log}(1+f^2)\big )~.\label{defA2}
\eea
$N_{\epsilon}^{(j)}(q,f_0,f)$ is the nominator in \eqref{rrr}, i.e.
\bea
N_{\epsilon}^{(j)}(q,f)&=&1-2\epsilon qf\mbox{cos}(\varphi(f)\pm \gamma_j)\label{Nj}\\[2mm]
&& +~ \sqrt{1-4\epsilon qf\mbox{cos}(\varphi\pm\gamma_j)-4\epsilon ^2q^2f^2\mbox{sin}^2(\varphi\pm\gamma_j)-4\epsilon ^2q^2}~.\nonumber
\eea
The straightforward evaluation of $A_{\epsilon}^{(j,2)}$ gives, using \eqref{Tj} and \eqref{gamma-cusp},
\bea
A_{\epsilon}^{(j,2)}&=&\Gamma_{\mbox{\tiny cusp}} (\theta )~\mbox{log}(2q\epsilon )~+~\frac{1-\cos\gamma_j}{\epsilon~q\sin ^2\gamma _j}~\Big (1-\mbox{log}(\frac{1-\cos\gamma_j}{\sin^2\gamma_j})\Big )\nonumber\\[2mm]
&&-\int _{f_0}^{\infty}df~\left (\sqrt{\frac{f^4+f^2}{f^4+f^2-E^2}}-1\right )~\mbox{log}(1+f^2)~+~2q~\mbox{log}(2qT_j )\nonumber\\[2mm ]
&&~ +~f_0~\mbox{log}(1+f_0^2)~+~2~\mbox{arctan}f_0-2f_0-\pi ~+~{\cal O}(\epsilon ~\mbox{log}\epsilon)~.\label{A2}
\eea
The bit more involved evaluation of $A_{\epsilon}^{(j,1)} $ is discussed in some detail in appendix C, with the result given in \eqref{A1final}. Making use of the definition of $\Gamma_{\mbox{\tiny cusp}} (\theta )$ in eq.\eqref{gamma-cusp} of appendix B
it appears as
\bea
A_{\epsilon}^{(j,1)}&=&\frac{2}{q\epsilon}~\left (\frac{\gamma_j}{2~\mbox{sin}\gamma_j}-\frac{1+\mbox{log}(1+\mbox{cos}\gamma_j)}{4~\mbox{cos}^2\frac{\gamma_j}{2}}\right )\nonumber\\[2mm]
&&~-~\Gamma_{\mbox{\tiny cusp}} (\theta )  ~\mbox{log}2~-~2q~\mbox{log}\Big(\frac{1-\mbox{cos}\gamma_j}{\mbox{sin}^2\gamma_j}\Big )~+~\mbox{{\tiny \cal O}}(1)~,\label{A1}
\eea
and then the sum \eqref{A1+A2} is
\bea
A_{\epsilon}^{(j)}&=&\Gamma_{\mbox{\tiny cusp}} (\theta )~\mbox{log}(q\epsilon )~+~
\frac{\gamma_j}{\epsilon~q~\sin\gamma_j}~+~f_0~\mbox{log}(1+f_0^2)~+~2~\mbox{arctan}f_0-2f_0-\pi\nonumber\\[2mm]
&&-~\int _{f_0}^{\infty}df~\left (\sqrt{\frac{f^4+f^2}{f^4+f^2-E^2}}-1\right )~\mbox{log}(1+f^2)~~+~\mbox{{\tiny \cal O}}(1)~.
\eea

To perform the  ambiguous separation into a divergent and finite part we have to introduce an RG-scale $\mu$. Then we get for the total area \eqref{AA1A2}
\beq
A_{\epsilon}~=~ 2 \Gamma_{\mbox{\tiny cusp}} (\theta )~\mbox{log}(\mu\epsilon )~+~\frac{l}{\epsilon}~+~{\cal O}(1)~.
\eeq
Here we have made use of appendix A, eq.\eqref{length} to confirm also for this explicit example the generic property, that the factor for the linear divergence
is given by the length of the contour. 

The renormalised area is ($Q=1/q$ distance between the cusps, see \eqref{Qq})
\bea
A_{\mbox{\tiny ren}}&=&- 2 \Gamma_{\mbox{\tiny cusp}} (\theta )~\mbox{log}\big (\mu Q \big )~+~2f_0~\mbox{log}(1+f_0^2)~+~4~\mbox{arctan}f_0-4f_0-2\pi\nonumber\\[2mm]
&&-~2~\int _{f_0}^{\infty}df~\left (\sqrt{\frac{f^4+f^2}{f^4+f^2-f_0^4-f_0^2}}-1\right )~\mbox{log}(1+f^2)~.
\eea
It has a remarkable structure. There is a conformally invariant contribution depending only on the cusp
angle $\theta$, via $f_0(\theta)$ given in \eqref{f0theta}. Dependence on other
data of our contour appear only via the first term, whose presence is enforced by the RG-ambiguity and which breaks conformal invariance. There this dependence comes via the  distance between the two
cusps $Q$, which by \eqref{Qq} and \eqref{pgammaDR} is given as a function of the radii of the two circles and the distance of their centers.  

Since the linear divergence is $l/\epsilon $, perhaps the most natural choice for the RG scheme is minimal
subtraction with respect to $l/\epsilon $. This corresponds to $\mu = 1/l$. Then $A_{\mbox{\tiny ren}}$ depends
on the cusp angle $\theta$ and the ratio  $Q/l$.

Closing this section, let us consider the limit in which our contour becomes a 
circle. Then one has $\theta=\pi$, i.e $f_0=0$ and $ Q=2R$. Since $ \Gamma_{\mbox{\tiny cusp}} (\pi )=0$, this gives
\beq
A_{\mbox{\tiny ren}}^{\mbox{\tiny circle}}~=~-2\pi~,
\eeq
in agreement with the literature \cite{Drukker:1999zq,Drukker:2000rr,Semenoff:2002kk}.
\section{Cusp anomalous dimension in the generic case}
Let us consider a closed contour in the  $(x_1,x_2)$-plane with a cusp, located at the origin, but smooth otherwise. At first we divide the related minimal surface in two
parts, depending on whether $\rho =\sqrt{x
_1^2+x_2^2}$ is smaller or larger then a certain value. This $\rho_0$ should be small, {\it but kept fixed for } $\epsilon \rightarrow 0$. The corresponding regularised area
is then given by
\beq
A_{\epsilon}~=~A_{\epsilon}^{\tiny \mbox{cusp}}(\rho _0)~+~A_{\epsilon}^{\tiny \mbox{smooth}}(\rho _0)~.\label{A-cusp-smooth}
\eeq
Due to the general result for smooth contours we know already
\beq
A_{\epsilon}^{\tiny \mbox{smooth}}(\rho _0)~=~\frac{l-l_{\rho_0}}{\epsilon}~+~{\cal O}(1)~,\label{A-smooth}
\eeq
where $l$ is the length of the total contour and $l_{\rho_0}$ that  of the cusp piece.
 
Let now the two curved legs of the cusp be parameterised in the vicinity of the origin by
\beq
x^{(j)}_1~=~\rho~\mbox{cos}\big (\phi^{(j)}(\rho )\big )~,~~~~x^{(j)}_2~=~\rho~\mbox{sin}\big (\phi^{(j)}(\rho )\big )~,~~~~j=1,2~.
\eeq
The cusp angle $\theta$ is then given by  
\beq
\theta~=~\phi_2(0)~-~\phi_1(0)~,
\eeq
and the $\rho \rightarrow 0$ limits of the curvatures of both legs are  
\beq 
k_j ~=~2c_j~= ~2~\frac{d\phi^{(j)}}{d\rho}\Big\vert _{\rho =0}~.\label{curvature}
\eeq
The length $l_{\rho_0}$ turns out to be
\beq
l_{\rho_0}~=~2\rho_0~+~\frac{1}{6}(c_1^2+c_2^2)\rho _0^3~+~{\cal O}(\rho _0^4)~.\label{l-cusp}
\eeq

For the evaluation of $A_{\epsilon}^{\tiny \mbox{cusp}}(\rho _0)$ we want to work with coordinates $\rho~,\varphi $ for the $(x_1,x_2)$-plane, adapted in a manner that still $\rho ^2=x_1^2+x_2^2$,
but that the lines of constant $\varphi =0 $ and $\varphi =\theta$, respectively, agree with the curved legs forming the cusp. The first requirement is realised by the structure
\beq
x_1~=~\rho~u(\rho,\varphi)~,~~~~x_2~=~\rho~\sqrt{1-u^2}~,
\eeq
and the second one means in addition
\beq
u(\rho,0)~=~\mbox{cos}\big (\phi^{(1)}(\rho)\big )~,~~~u(\rho,\theta)~=~\mbox{cos}\big (\phi^{(2)}(\rho)\big )~.
\eeq
The additional $AdS$-coordinate $r$ we parameterise by 
\beq
r~=~\rho~F(\rho,\varphi)~,
\eeq
with the boundary condition
\beq
F(\rho,0)~=~F(\rho,\theta)~=~0~.
\eeq
Then the regularised area of the cusp piece is given by 
\beq
A_{\epsilon}^{\tiny \mbox{cusp}}(\rho _0)~=~\int_{r>\epsilon,~\rho<\rho_0}L(\rho,\varphi)~d\rho d\varphi~,\label{A-eps}
\eeq
with \footnote{A dot means a derivative w.r.t. $\rho$, a prime w.r.t. $\varphi $.}
\beq
L(\rho,\varphi)=\sqrt{\frac{F'^2\big (1-u^2+\rho ^2\dot{u}^2\big )+\big (1+(F+\rho \dot{F})^2\big )u'^2- 2\rho F'\big (F + \rho \dot{F}\big ) \dot{u}u'}{\rho ^2F^4(1-u^2)}}~.\label{lag}
\eeq

We will look for perturbative solutions of the related equation of motion  \footnote{
$~$i.e. the Euler-Lagrange equation obtained by varying $F$. Since $u$ is a building block for the 
\\$~~~~~~~$specification of the coordinate system, it is kept fixed in this process.} in the form
\beq
F(\rho,\varphi)~=~F_1(\varphi)~+~\rho ~F_2(\varphi)~+~\dots~,\label{F-exp}
\eeq
with the boundary conditions 
\beq
F_n(0)~=~F_n(\theta)~=~0~,~~~~n=1,2,\dots ~~.\label{bc-Fn}
\eeq
As a more concrete form for $u$  we now take
\beq
u(\rho,\varphi)~=~\mbox{cos}\left (\frac{\rho ~s(\varphi)+\varphi}{\theta}~\big (\phi_2(\rho )-\phi_1(\rho)\big )~+~\phi_1(\rho)\right )~,\label{u-rho-phi}
\eeq
where the function $s(\varphi)$ has to be chosen with the behaviour
\bea
s(\varphi)&=&a_1~\varphi ^{2/3}~+~\dots ~~,~~~~~~~~~~\varphi \rightarrow 0 ~,\nonumber \\
s(\varphi)&=&a_2~(\theta -\varphi ) ^{2/3}~+~\dots ~~,~~~\varphi \rightarrow \theta~.
\label{s-asympt}
\eea
The $a_j$ are constants to be fixed later.

Let us stop for a moment in order to comment on the need for this peculiar $s(\varphi)$. Naively one would of course start
with $s(\varphi)$ put to zero in \eqref{u-rho-phi}. But then, as we will see below, the boundary condition \eqref{bc-Fn} 
for $F_2$ cannot be fulfilled. For more intuitive arguments for this ansatz, beyond this a posteriori justification, see appendix D. There one also finds in fig.\ref{new-coordinate-plot} a visualisation of the difference of coordinates with and without $s(\varphi )$.

The variational derivative of \eqref{lag} with respect to $F$ appears as a quotient.
We take as the equation of motion the vanishing of the nominator, insert \eqref{u-rho-phi} and expand the result with respect to $\rho$. Then the condition, that in this 
expansion the coefficients of the leading and nextleading term vanish, leads to
\beq
2 (F_1')^2 + (1 + F_1^2) (2 + F_1^2 + F_1 F_1'')~=~0 \label{leading-eom}
\eeq
and
\beq
F_2''(\varphi)+ G_1(\varphi)F_2'(\varphi) + G(\varphi) F_2(\varphi) + M(\varphi)~=~0~,\label{nextleading-eom}
\eeq
in which  \footnote{A term, vanishing due to the leading order equation \eqref{leading-eom} for $F_1$, has been omitted already in 
\\$~~~~~~~$the expression for $M$. $c_j$ defined in \eqref{curvature}.}
\bea
M(\varphi)&=&\frac{-1}{\theta(F_1+F_1^3)}~\Big \{~\theta (F_1+F_1^3)F_1'~s''(\varphi)\nonumber\\[2mm]
 &&~~~~~+\big (c_1(\theta-\varphi)+c_2\varphi +\theta s(\varphi)\big )\big (2F_1(F_1')^3+F_1F_1'(7+3F_1^2)\big ) \label{M}\\[2mm]
&&~~~~~~~+\big (c_1-c_2-\theta s'(\varphi)\big )(1+F_1^2)\big (6+3F_1^2+2(F_1')^2+F_1F_1''\big )\Big \}~,\nonumber\\[5mm]
G(\varphi )&=&\frac{13F_1+7F_1^3+2F_1(F_1')^2+(1+5F_1^2)F_1''}{F_1+F_1^3}~,\label{G}\\[5mm]
G_1(\varphi)&=&\frac{2(2-F_1^2)~F_1'}{F_1+F_1^3}~.\label{G1}
\eea

As expected, the leading order equation \eqref{leading-eom} depends on $F_1$ only. Moreover, it equals that for the full $F$ in the case with straight
legs, see \eqref{eq-F}. Therefore we know from appendix B $F_1(\varphi)=F_1(\theta -\varphi)$ and in particular from eq.\eqref{F-expansion}
\beq
F_1(\varphi)~=~a \varphi^{1/3}~+~\frac{a^3}{5}\varphi ~+~{\cal O}(\varphi ^{5/3})~,~~~~a=(3/E)^{1/3}~.\label{asF1}
\eeq
Since our concern are divergent contributions to the regularised area 
\beq
A_{\epsilon}~=~\int _{\rho F(\rho,\varphi)>\epsilon}L(\rho,\varphi)d\rho d\varphi
\label{area2}
\eeq
for $\epsilon\rightarrow 0$, we also have to control the behaviour of $F_2$ near the $AdS$ boundary. We will do this for $\varphi\rightarrow 0$, the case $\varphi\rightarrow \theta$ looks similarly.  
From \eqref{s-asympt} and \eqref{asF1} we get for $\varphi\rightarrow 0$
\newpage
\bea
M(\varphi)&=&\frac{2(aa_1-a^3c_1)}{27\varphi ^2}~+~\frac{8(2a^3a_1-a^5c_1)}{135\varphi^{4/3}}~+~{\cal O}(1/\varphi )~,\label{M-asympt}\\
G(\varphi)&=&-\frac{2}{9\varphi ^2}~-~\frac{28 a^2}{45 \varphi ^{4/3}}~+~{\cal O}(\varphi^{-2/3})~,\label{G-asympt}\\
G_1(\varphi)&=&\frac{4}{3\varphi}~-~\frac{22a^2}{15\varphi^{1/3}}~+~{\cal O}(\varphi^{1/3})~.\label{G1-asympt} 
\eea
Due to this type of singular behaviour of its coefficient functions, the differential equation for $F_2(\varphi)$ is singular at $\varphi =0$. Since $\varphi ^2 M~,~\varphi ^2 G$ and $\varphi G_1$ are analytic in $x=\varphi ^{1/3}$, we have a regular singular point, and $F_2$ can be represented as a Frobenius series in $x$. 
Inserting the leading plus nextleading terms for $G$, $G_1$ and $M$ from \eqref{M-asympt}-\eqref{G1-asympt}  into \eqref{nextleading-eom} one gets ($B_1,B_2$ integration constants)
\bea
 F_2(\varphi)&=&\frac{1}{3}(a a_1-a^3 c_1)-\frac{( 4 a^3 a_1 
- 2 a^5 c_1)}{15} ~\varphi^{2/3} + {\cal O}(\varphi)\label{asF2}\\[2mm]
&&~~~~~~~~~+ B_1 \big (\varphi^
{-2/3}+\frac{8}{5}a^2+{\cal O}(\varphi ^{2/3})\big )+B_2\big ( \varphi^{1/3}+a^2\varphi+ {\cal O}( \varphi^{5/3})\big )~.\nonumber
\eea

This expression contains three so far unfixed parameters of quite different origin. $a_1$ specifies the choice of the coordinate system and $B_1$ and $B_2$ are the free parameters in the general solution of the differential equation.  
Now the boundary condition $F_2(0)=0$ requires first of all $B_1=0$ in order to prevent a divergence. But then there is still the first term on the r.h.s. of \eqref{asF2} obstructing the boundary condition. It vanishes only for the choice 
\beq
a_1~=~a^2c_1 ~.\label{A1c1}
\eeq
As announced above, we see that without $s(\varphi)$ in \eqref{u-rho-phi} the boundary condition for  $F_2$ cannot be fulfilled, if the corresponding leg of the cusp
is curved \footnote{We have given already arguments for $2/3$ as  the exponent in \eqref{s-asympt}. As an interesting aside one can perform the analysis 
with
a generic exponent $b$. Then the leading term for $M(\varphi)$ in 
\eqref{M-asympt} turns out to be  $\propto \varphi ^{b-8/3}$ for $b<2/3$. It is $\propto 1/\varphi ^2$ for $b>2/3$, but with a nonzero coefficient (as soon as $c_1\neq 0$), independent of the choice of $a_1$.  }. Fixing $B_2$ (and $a_2$ in \eqref{s-asympt}) is  a matter of the boundary condition on the second leg of the cusp.

To analyse the regularised area \eqref{area2}, we expand $L(\rho,\varphi)$ with respect to $\rho$
\beq
L(\rho,\varphi)~=~\frac{1}{\rho}~L_1(\varphi)~+~L_2(\varphi )~+~{\cal O}(\rho)~.
\eeq
Then we get with \eqref{lag},\eqref{F-exp},\eqref{u-rho-phi}
\beq
L_1(\varphi )~=~\frac{\sqrt{1+F_1^2+(F_1')^2}}{F_1^2}~,
\eeq
\bea
L_2(\varphi)&=&\frac{1}{\theta ~F_1^3\sqrt{1+F_1^2+(F_1')^2}}~\Big \{
\big (c_2 -c_1+\theta s'(\varphi)\big )\big (F_1+F_1^3\big )\\[2mm]
&&-\big (c_1(\theta-\varphi)+c_2\varphi +\theta s(\varphi)\big )F_1^2F_1'-2\theta \big(1+(F_1')^2\big )F_2+\theta F_1F_1'F_2'\nonumber
\Big \}~.
\eea
Due to the asymptotics \eqref{asF1} this means for $\varphi \rightarrow 0$
\beq
L_1(\varphi )~=~\frac{1}{3a\varphi ^{4/3}}~+~\frac{a}{15\varphi ^{2/3}}~+~{\cal O}(1)
~,
\eeq
and via \eqref{asF1},\eqref{asF2} and \eqref{A1c1}
\beq
L_2(\varphi )~=~-B_2\Big (\frac{1}{3a^2\varphi ^{4/3}}~+{\cal O}(\frac{1}{\varphi^{2/3}})\Big )~+~
\frac{c_1}{a\varphi ^{1/3}}~+~{\cal O}(1)~.\label{L2-as}
\eeq  
If $B_2\neq 0$, the integral of $L_2(\varphi )$ is still divergent. The further analysis would be simpler
if we could put $B_2=0$. But, as mentioned already and elaborated in more detail in appendix E, fixing $B_2$ is a matter
of the boundary condition on the second leg ($\varphi\rightarrow\theta $), and for generic situations we have to live with
$B_2\neq 0$.\\

The divergences in \eqref{A-eps} arise from the neighbourhood of $\rho =0$,
$\varphi =0$ and $\varphi =\theta$. To keep them under control with the above
estimates, we write
\beq
A_{\epsilon}^{\tiny \mbox{cusp}}(\rho _0)~=~A_{\epsilon}^{\tiny \mbox{c},0}(\rho _0)~+~A_{\epsilon}^{\tiny \mbox{c},\theta}(\rho _0)\label{cusp-0-theta}
\eeq
and understand the first term on the r.h.s with the $\varphi$-integration
over the interval  $(0,\theta/2$) and the second one over $(\theta/2,\theta)$. 
Then one has for the second order contribution to $A_{\epsilon}^{\tiny \mbox{c},0}(\rho _0)$
\bea
~A_{\epsilon,2}^{\tiny \mbox{c},0}(\rho _0)&=&\int _{\rho <\rho_0,~~\rho F_1+\rho ^2F_2+{\cal O}(\rho^3)>\epsilon}~ L_2(\varphi)~d\rho d\varphi\nonumber\\[3mm]
&=&-\frac{B_2}{3a^2}\int_{\rho_{min}(\epsilon)}^{\rho _0}d\rho\int ^{\theta/2}_{(a\rho+B_2\rho ^2)\varphi^{1/3}+{\cal O}(\rho^3)=\epsilon}~d\varphi ~\varphi^{-4/3}~+~{\cal O}(1)\nonumber\\[3mm]
&=&-\frac{1}{\epsilon}~\Big (\frac{B_2}{2a}~\rho_0^2+\frac{B_2^2}{3a^2}~\rho_0^3+{\cal O}(\rho _0^4)\Big )~+~{\cal O}(1)~.\label{cusp-2}
\eea
Use has been made of \eqref{L2-as} and $\rho_{min}={\cal O}(\epsilon)$.

The integrand of the leading order contribution $A_{\epsilon,1}^{\tiny \mbox{c},0}(\rho _0)$ is the same as for
the case with a cusp between straight legs. However, the integration region
depends also on $F_2$. From \eqref{Aeps1} in appendix F we can take
\bea
~A_{\epsilon,1}^{\tiny \mbox{c},0}(\rho _0)&=&\int _{\rho <\rho_0,~~\rho F_1+\rho ^2F_2+{\cal O}(\rho^3)>\epsilon}~ \frac{1}{\rho}~L_1(\varphi)~d\rho d\varphi\nonumber\\[3mm]
&=&\frac{1}{\epsilon}~\Big( \rho_0+\frac{B_2}{2a}\rho _0^2+{\cal O}(\rho_0^3)\Big )~+~\frac{1}{2}~\Gamma _{\mbox{\tiny cusp}}(\theta)~\mbox{log}\epsilon ~+~{\cal O}(1)~.\label{cusp-1}
\eea
Note that the $1/\epsilon $ terms proportional to $\rho_0^2$ in the sum of \eqref{cusp-2} and \eqref{cusp-1} cancel.\\

Repeating the same estimates \footnote{Then $B_2$ has to be replaced by $\hat B_1:=v_1(\theta)$, see appendix E. } for $A_{\epsilon}^{\tiny \mbox{c},\theta}(\rho )$ we get with \eqref{cusp-0-theta}-\eqref{cusp-1}
\beq
A_{\epsilon}^{\tiny \mbox{cusp}}(\rho _0)~=~\frac{1}{\epsilon}~\big(2 \rho_0+{\cal O}(\rho _0^3)\big )~+~\Gamma _{\mbox{\tiny cusp}}(\theta)~\mbox{log}\epsilon ~+~{\cal O}(1)~,\label{A-cusp}
\eeq
and finally with \eqref{A-cusp-smooth},\eqref{A-smooth},\eqref{l-cusp} and \eqref{A-cusp}
\beq
A_{\epsilon}~=~\frac{l}{\epsilon}~+~\Gamma_{\mbox{\tiny cusp}}(\theta)~\mbox{log}\epsilon~+~{\cal O}(1)~.
\eeq

Why we are sure that the ${\cal O}(\rho _0^3)$ terms present in $A_{\epsilon}^{\tiny \mbox{cusp}}(\rho _0)$ and $A_{\epsilon}^{\tiny \mbox{smooth}}(\rho _0)$ cancel each other? If we a priori assume, that $A_{\epsilon}$ has an asymptotic expansion with just an $1/\epsilon$ and log$\epsilon $ singular term, then it follows
due to the uniqueness of asymptotic expansions and the absence of any $\rho_0$-dependence in $A_{\epsilon}$ itself.  Without this assumption, we can rely on the fact that
the $\epsilon$-expansions of both $A_{\epsilon}^{\tiny \mbox{cusp}}(\rho _0)$ and $A_{\epsilon}^{\tiny \mbox{smooth}}(\rho _0)$ are valid uniformly in a certain interval for $\rho _0$. Then the difference of the resulting expansion for $A_{\epsilon}$ at two different values of $\rho_0$ is an expansion of zero with necessarily zero coefficients, and this just proves the stated $\rho_0$-independence.  

To complete the proof, we still have to comment on the effect of $F_3$ and higher orders in \eqref{F-exp}. 
For this purpose it is not necessary to continue order by order the tedious derivation of the corresponding differential equations and  their asymptotic estimates for $\varphi\rightarrow 0$. Instead we argue, that $F(\rho,\varphi)$ 
as a whole has to behave like $ \varphi^{1/3}$ for all fixed $\rho>0$, to fit the asymptotic information contained in
\eqref{surface-asympt}. Then \eqref{lag} implies $L(\rho,\varphi)\propto \varphi^{-4/3}$ and consequently $L_n(\varphi)\propto \varphi^{-4/3}$ for all $n$. The contribution of $L_n$ to
$A_{\epsilon}^{\tiny \mbox{cusp}}(\rho _0)$ is
$$\int _{\rho <\rho_0,~\rho F(\rho,\varphi)>\epsilon}\rho^{n-2}L_nd\rho d\varphi~. $$
Obviously, for $n\geq 3$ it does not contribute to the logarithmic divergence. By analogous
arguments as above for the $L_2$ contribution, its $\rho_0$
dependent contribution to the $1/\epsilon$ term just cancels the corresponding $\rho _0$ dependence in $A_{\epsilon}^{\tiny \mbox{smooth}}(\rho _0)$.
\section{Conclusions}
Concerning our calculation of the regularised area $A_{\epsilon}$ for a minimal
surface related to Wilson loops for contours formed by segments of intersecting 
circles, two points should be emphasised. The usual calculation for a cusp formed
by two straight half-lines needs both an UV cutoff $r>\epsilon$ as well as
an IR cutoff $L$ with the result given in \eqref{straight-case}. There $\Gamma_{\mbox{\tiny cusp}}$ appears both as the factor in front of the logarithmic UV divergence
due to the cusp as well as the factor (with a minus sign) of the logarithmic
IR divergence due to the infinite extension of the half-lines. 

Our calculation in section 2 is the first example of an explicit calculation
of  the minimal surface and its area $A_{\epsilon}$ for a curved contour with 
cusps closed in a finite domain
 \footnote{The only other finite cusped contours with explicitly known surface are the null tetragon  \cite{alday-malda} and a certain degenerated null hexagon \cite{Sakai:2009ut}, both relevant for scattering amplitudes. But there the 
contour is straightly, away from the cusps. For progress in surface construction
related to smooth boundary curves see e.g. \cite{Irrgang:2015txa} and references therein.}. 
It has two cusps
with equal angle and needs no IR regularisation. As the coefficient of the linear
UV divergence we get the length of the contour as required by the general
theory for smooth contours. For the logarithmic divergence we get
the expected factor $ 2~\Gamma_{\mbox{\tiny cusp}}$. The renormalised area
in a most  natural minimal subtraction scheme turns out to be a function
of the cusp angle and the ratio of the distance of the cusps to the length of the contour.

Furthermore, with the just discussed case we have at hand an example, which confirms the expectation, that the cusp anomalous dimension depends only on the cusp angle, also
in cases where the legs of the cusp are curved. This has then been verified in section 3 for the larger class of generic cusped contours in a Euclidean plane. 
One by-product of the technique developed for this purpose could be
the use of numerical solutions of the differential equation
in appendix E for control over the nextleading contribution in a perturbative construction of the surface in the vicinity of the cusp. 

To cover full generality beyond planar contours one has to repeat our analysis,
but now with additional functions of $\rho$ and $\varphi$, which describe the
extension of the surface in the directions of $x_0$ and $x_3$.\\ [20mm]
 \noindent
 {\bf Acknowledgement:}\\[2mm]
 I would like to thank Danilo Diaz, Nadav Drukker, Hagen M\"unkler and in 
particular George Jorjadze for useful discussions.
\newpage
\section*{Appendix A}
Here we want to provide the elementary geometric formulas related to
the conformal mapping of two crossing circles to two crossing straight lines.

Under an inversion
\beq
y_{\mu}~=~\frac{x_{\mu}}{x^2}~~~~~\Longleftrightarrow ~~~~~x_{\mu}~=~\frac{y_{\mu}}{y^2}\label{inversion}
\eeq  
two straight lines in the $(x_1,x_2)$-plane, crossing the $x_1$-axis at $x_1=q>0$ with  angles $\gamma_1 $ and $\gamma_2$, respectively,
are mapped to two circles in the $(y_1,y_2)$-plane, crossing at  the origin
and at $(1/q,0)$.  Hence the distance between the two crossing points is
\beq
Q~=~\frac{1}{q}~.\label{Qq}
\eeq
The radii of the circles are
\beq
R_j~=~\frac{1}{2q~\vert \mbox{sin}\gamma_j\vert}~,\label{Rpgamma}
\eeq   
and the centers of the circles are located at
$$
\frac{1}{2q}~(1,\mbox{cot}\gamma_j)~,~~~~~j=1,2
$$
respectively.
The distance between the two centers is
\beq
D~=~\frac{1}{2q}~\vert \mbox{cot}\gamma_1-\mbox{cot}\gamma_2\vert~.\label{Dpgamma}
\eeq

Inverting \eqref{Rpgamma},\eqref{Dpgamma}, one gets $q$ and the $\gamma_j$ in terms of $D$ and the $R_j$
\bea
q&=&\frac{D}{\sqrt{2D^2(R_1^2+R_2^2)-D^4-(R_1^2-R_2^2)^2}}\nonumber\\[5mm]
\vert \mbox{sin}\gamma _j\vert &=&\frac{\sqrt{ 2D^2(R_1^2+R_2^2)-D^4-(R_1^2-R_2^2)^2}}{2DR_j}~.\label{pgammaDR}
\eea
With the convention $R_1\geq R_2$, these formulas are valid in the full range
\beq
R_1-R_2~\leq~D~\leq~R_1+R_2~.\label{R1R2D}
\eeq
Obviously, beyond these bounds there is no crossing. Let us also note that
\eqref{pgammaDR} and \eqref{R1R2D} imply
\beq 
\vert\mbox{sin}\gamma_1\vert\leq\frac{R_2}{R_1}~,~~~\vert\mbox{sin}\gamma_2\vert\leq 1~.
\eeq
The length of the contour, formed by the two segments of the crossing circles is
\beq
l~=~\frac{1}{q}\Big ( \frac{\gamma_1}{\sin\gamma_1}~+~\frac{\gamma_2}{\sin\gamma_2}\Big )~.\label{length}
\eeq
 
\begin{figure}[h!]
 \centering
 \includegraphics[width=9cm]{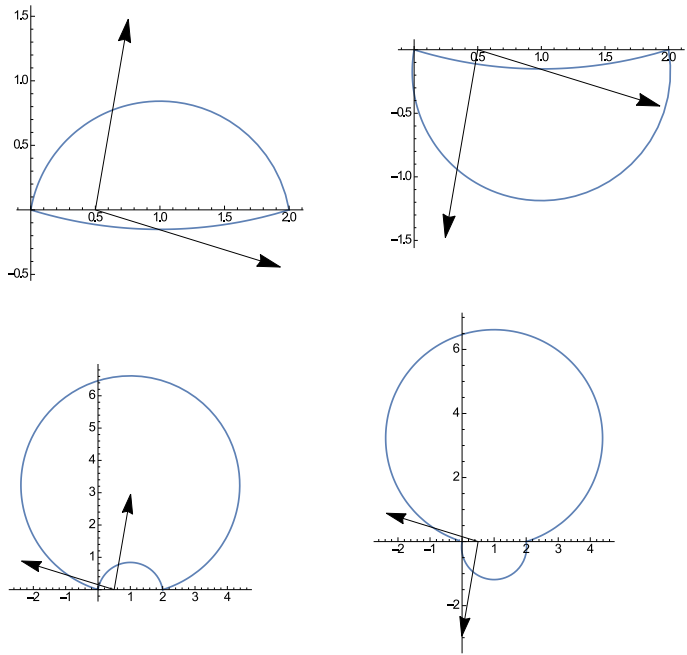} 
\caption{\it The four closed contours (blue) with cusps, formed out of circular segments, as images of rays (black), starting at (q,0) with q=0.5. Note the difference in scale between the first and second line.} 
\label{4-cusped-contours}
\end{figure}

Let us now consider two rays (i.e. two halfs of the straight lines of the previous discussion), starting on the $x_1$-axis at $(q,0)$. Their images under 
\eqref{inversion} form the boundary of a compact region with two cusps. For this we have four possibilities, two of these regions are convex and two not convex, see fig.\ref{4-cusped-contours}.
\section*{Appendix B}
In this appendix we review the calculation for a cusp with straight legs based on \cite{Drukker:1999zq}, \cite{Kruczenski:2002fb}. Furthermore, we
present an alternative representation of the cusp anomalous dimension in terms of a  hypergeometric function and
comment on a parameterisation of the problem, more convenient for our discussion in the main text. 

Due to the dilatation symmetry of a cusp located at the origin between two {\it straight} legs extending {\it to infinity},  one can describe
the $r$-coordinate of the string surface by the ansatz
\beq
r~=~\frac{\rho}{f(\varphi)}~,
\eeq
where  $\rho ,\varphi$ are polar coordinates in the $(x_1,x_2)$-plane of the $AdS$ boundary at $r=0$, i.e.
\beq
x_1~=~\rho~\mbox{cos}~\varphi~,~~~~~x_2~=~\rho ~\mbox{sin}~\varphi ~.
\eeq
Then the determinant of the induced metric on the surface turns out to be
\beq
h~=~\frac{f^4+f^2+(f')^2}{\rho ^2}~,\label{ind-metric-det}
\eeq
and, in virtue of this factorisation of the $\rho $ and $\varphi $ dependence, the minimal surface condition (equation of motion) is an ordinary differential equation for $f(\varphi )$ with boundary conditions $f(0)=f(\theta)=\infty$ ($\theta $ cusp angle) 
\beq
f''(f^4+f^2)-2(f')^2(f+2f^3)-f^3(1+f^2)(1+2f^2)~=~0~.\label{eom-f}
\eeq
Instead of solving this second order equation directly, it is more convenient
to use the conservation law related to the lack of any  explicit $\varphi$-dependence in \eqref{ind-metric-det}
\beq
E~=~f_0\sqrt{1+f_0^2}~=~\frac{f^4+f^2}{\sqrt{f^4+f^2+(f')^2}}~,\label{E-f}
\eeq
with $f_0=f(\theta /2)$ denoting the minimal value of $f$ in $\varphi\in (0,\theta)$. Now integration yields
\beq
\varphi ~=~E~\int _f^{\infty}\frac{df}{\sqrt{(f^4+f^2)^2-E^2(f^4+f^2)}}~.\label{phi-f}
\eeq
This holds for $0<\varphi<\theta /2$, and furthermore one has $f(\varphi)=f(\theta-\varphi)$. 

In particular the equation
\beq
\theta ~=~2E~\int _{f_0}^{\infty}\frac{df}{\sqrt{(f^4+f^2)^2-E^2(f^4+f^2)}}~\label{f0theta}
\eeq
fixes the relation between $f_0$ and the cusp angle $\theta $. The function $\theta(f_0)$ is
monotonically decreasing, $\theta(0)=\pi,~~\theta(\infty)=0$. 
The estimate of this equation and \eqref{E-f} for large $f_0$ yields
\beq
\theta~=~\frac{2\pi ^{1/2} \Gamma(\frac{3}{4})}{\Gamma(\frac{1}{4})}~f_0^{-1}~+~{\cal O}(f_0^{-3})~.
\eeq

Expanding the integrand in eq.\eqref{phi-f} for large $f$ (but fixed $f_0$) one gets
\beq
\frac{\varphi}{E}~=~\frac{1}{3}f^{-3}-\frac{1}{5}f^{-5}+\frac{1}{7}(1+\frac{E^2}{2})f^{-7}+{\cal O}(f^{-9})~.\label{phi-f-as}
\eeq
The  inversion of \eqref{phi-f-as} is
\beq
\frac{1}{f(\varphi)}~=~\Big (\frac{3\varphi}{E}\Big )^{1/3}~+~\frac{1}{5}\cdot\frac{3\varphi}{E}~+~\frac{6-25 E^2}{350}\cdot \Big (\frac{3\varphi}{E}\Big )^{5/3}~+~{\cal O}(\varphi ^{7/3})~.\label{F-expansion}
\eeq
For use in the main text we note that
\beq
F(\varphi):=\frac{1}{f(\varphi)}~\label{f-F}
\eeq
defines a function which has an ordinary Taylor expansion in $x=\varphi ^{1/3}$ around $x=0$. Expressing the equation of motion \eqref{eom-f} in terms of $F(\varphi)$ it looks like
\beq
2~(F')^2~+~(1+F^2)~(2+F^2+FF'')~=~0~.\label{eq-F}
\eeq

For  the regularised area, defined with the cutoffs $r=\rho /f(\varphi)>\epsilon$ and $\rho <L$, one gets
\bea
A_{\epsilon,L}&=&\int d\rho~ d\varphi~ \frac{\sqrt{f^4+f^2+(f')^2}}{\rho}\nonumber\\
&=&\frac{2L}{\epsilon}~+~\Gamma_{\mbox{\tiny cusp}}(\theta)~\mbox{log}\frac{\epsilon}{L}~+~
A_0(\theta) ~+~\dots ~.\label{straight-case}
\eea
The dots denote terms vanishing for $\epsilon \rightarrow 0$ and
\beq
\Gamma_{\mbox{\tiny cusp}} (\theta )~=~2f_0~-~2\int _{f_0}^{\infty}\left (\sqrt{\frac{f^4+f^2}{f^4+f^2-E^2}}-1\right )df~,\label{gamma-cusp}
\eeq
\beq
A_0(\theta )~=~2f_0~(\mbox{log}f_0~-~1)~-~2\int _{f_0}^{\infty}\mbox{log}f\left (\sqrt{\frac{f^4+f^2}{f^4+f^2-E^2}}-1\right )df~.
\eeq
The substitution $f^2=f_0^2+z^2$ yields  \cite{Kruczenski:2002fb}
\beq
\Gamma_{\mbox{\tiny cusp}} (\theta )~=~\int _{-\infty}^{\infty}\left (1-\sqrt{\frac{1+z^2+f_0^2}{1+z^2+2f_0^2}}~\right )dz~.
\eeq
Based on this integral we found a closed expression in terms of a hypergeometric function
\beq
\Gamma_{\mbox{\tiny cusp}} (\theta )~=~\frac{\pi}{2}\frac{f_0^2}{\sqrt{1+f_0^2}}~~ _2F_1\Big (\frac{1}{2},\frac{3}{2},2,\frac{-f_0^2}{1+f_0^2}\Big )~.
\eeq
This implies that $\Gamma_{\mbox{\tiny cusp}}$ for $\theta\rightarrow\pi$, i.e. $f_0\rightarrow 0$ goes to zero and for $\theta\rightarrow 0$ i.e. $f_0\rightarrow\infty$ grows linearly in $f_0$ with a factor $\frac{\sqrt{\pi}~\Gamma(3/4)}{2~\Gamma(5/4)}$.
\begin{figure}[h!]
 \centering
 \includegraphics[width=9cm]{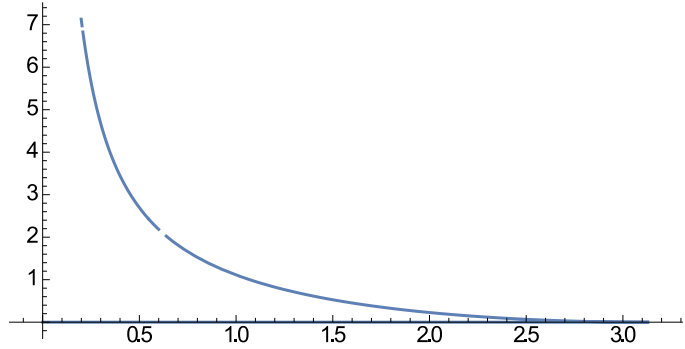} 
\caption{\it $\Gamma_{\mbox{\tiny cusp}}$ as a function of $\theta$, obtained with $f_0$ as parameter in ParametricPlot. } 
\label{gamma-cusp-as-fct-theta.eps}
\end{figure}
\section*{Appendix C}
This appendix is devoted to the $\epsilon\rightarrow 0$ asymptotics of $A_{\epsilon}^{(j,1)}$, as defined in \eqref{defA1}. We start with
\bea
A_{\epsilon}^{(j,1)}&=&2~\int _{f_0}^{f_{\epsilon}^{(j)}} df~\left (\sqrt{\frac{f^4+f^2}{f^4+f^2-E^2}}-1\right )~\mbox{log}N_{\epsilon}^{(j)}(q,f) \nonumber\\[2mm]
&&+~2~\int _{f_0}^{f_{\epsilon}^{(j)}} df~\mbox{log}N_{\epsilon}^{(j)}(q,f)~.
\eea
In the first term of the above equation the factor in big brackets is ${\cal O}(1/f^4)$ for large $f$. 
Let us split the integration region into the intervals $(f_0,T_j/\epsilon^{1/2})$ and $(T_j/\epsilon^{1/2},f^{(j)}_{\epsilon})$. Then in the first interval $\mbox {log}N_{\epsilon}^{(j)}(q,f)$ approaches $\mbox{log}2$. This is no longer true in the second interval, but due to the fast vanishing of the  factor in big brackets the corresponding
integral drops out for $\epsilon\rightarrow 0$. This implies
\bea
A_{\epsilon}^{(j,1)}&=&2~\mbox{log}2~\int_{f_0}^{\infty}df~\left (\sqrt{\frac{f^4+f^2}{f^4+f^2-E^2}}-1\right )\nonumber\\[2mm]
&&+~2~\int _{f_0}^{f_{\epsilon}^{(j)}} df~\mbox{log}N_{\epsilon}^{(j)}(q,f)~+~\mbox{{\tiny \cal O}}(1)~.\label{Aj1}
\eea
Denoting the second line of the last equation \footnote{For notational convenience we drop here the index $j$.}
by $J_{\epsilon}$ and using the same splitting of the integration interval, we get after the substitution $x=qf\epsilon$ with \eqref{Tj} and \eqref{Nj}
\beq
J_{\epsilon}=~J_{\epsilon}^{\mbox{\tiny lower}}~+~J_{\epsilon}^{\mbox{\tiny upper}}~=
~\frac{2}{q\epsilon}~\left (\int_{q\epsilon f_0}^{q T \epsilon^{1/2}}~+~\int^{qT-\epsilon q^2}_
{qT\epsilon ^{1/2}}\right )~\mbox{log}N_{\epsilon}(q,\frac{x}{q\epsilon})~dx~.\label{J}
\eeq
Then in $J_{\epsilon}^{\mbox{\tiny lower}}$ the integration variable $x$ is small, allowing an expansion of
$\mbox{log}N_{\epsilon}(q,\frac{x}{q\epsilon})$ in terms of $x$. This yields
\beq
J_{\epsilon}^{\mbox{\tiny lower}}~=~\frac{2T}{\epsilon^{1/2}}~\mbox{log}2~-~2f_0~\mbox{log}2~-~\frac{4}{q\epsilon}~\int _{q\epsilon f_0}^{qT\epsilon ^{1/2}}x~\mbox{cos}\big (\varphi\big (\frac{x}{q\epsilon}\big )\pm \gamma \big )~dx~+~\mbox{{\tiny \cal O}}(1)~.\label{lower}
\eeq
In $J_{\epsilon}^{\mbox{\tiny upper}}$ the argument of $\varphi (f)$, i.e. $f=x/(q\epsilon)$, tends to infinity
where $\varphi =0.$ This approach to zero is fast enough to justify
\beq
J_{\epsilon}^{\mbox{\tiny upper}}~=~\frac{2}{q\epsilon}~\int ^{qT-\epsilon q^2}_
{qT\epsilon ^{1/2}}dx ~\mbox{log}\Big (1-2x~\mbox{cos}\gamma+\sqrt{1-4x~\mbox{cos}\gamma-4x^2\mbox{sin}^2\gamma}~\Big )~+~\mbox{{\tiny \cal O}}(1)~.
\eeq
Writing this as an integral over $(0,qT)$ minus integrals over $(0,qT\epsilon^{1/2})$ and over $(qT-\epsilon q^2,qT)$ we get
\bea
J_{\epsilon}^{\mbox{\tiny upper}}&=&\frac{2}{q\epsilon}~\int ^{qT}_
{0}dx ~\mbox{log}\Big (1-2x~\mbox{cos}\gamma+\sqrt{1-4x~\mbox{cos}\gamma-4x^2\mbox{sin}^2\gamma}~\Big )\nonumber\\[2mm]
&&~-~2q~\mbox{log}\Big (1-2qT~\mbox{cos}\gamma+\sqrt{1-4qT~\mbox{cos}\gamma-4q^2T^2\mbox{sin}^2\gamma}~\Big )\nonumber\\[2mm]
&&~-\frac{2T}{\epsilon^{1/2}}~\mbox{log}2~+~\frac{4}{q\epsilon}~\int _{0}^{qT\epsilon ^{1/2}}x~\mbox{cos}\gamma  ~dx~+~\mbox{{\tiny \cal O}}(1)~.\label{upper}
\eea
In the sum $J_{\epsilon}^{\mbox{\tiny lower}}+J_{\epsilon}^{\mbox{\tiny upper}}$ the terms $\propto 1/\epsilon^{1/2}$ cancel. A further cancellation (up to vanishing terms)
 takes place for the  last terms in \eqref{lower} and \eqref{upper}. Then, using the integral
\bea
~&&\int ^{\frac{1-\mbox{\tiny cos}\gamma}{2~\mbox{\tiny sin}^2\gamma}}_0dx~\mbox{log}\Big (1-2x~\mbox{cos}\gamma+\sqrt{1-4x~\mbox{cos}\gamma-4x^2\mbox{sin}^2\gamma}~\Big )\nonumber\\[2mm]
&&~~~~~~~~~~~~~~~~~~~~~=~\frac{\gamma}{2~\mbox{sin}\gamma}~-~\frac{1+\mbox{log}(1+\mbox{cos}\gamma)}{4~\mbox{cos}^2\frac{\gamma}{2}}~,
\eea
and inserting \eqref{lower}, \eqref{upper} and \eqref{J} into \eqref{Aj1} we get \footnote{Reintroducing the index $j$.} with $T_j$ given by
\eqref{Tj} 
\bea
A_{\epsilon}^{(j,1)}&=&\frac{2}{q\epsilon}~\left (\frac{\gamma_j}{2~\mbox{sin}\gamma_j}-\frac{1+\mbox{log}(1+\mbox{cos}\gamma_j)}{4~\mbox{cos}^2\frac{\gamma_j}{2}}\right )\nonumber\\[2mm]
&&~+2~\mbox{log}2~\left (\int_{f_0}^{\infty}df~\left (\sqrt{\frac{f^4+f^2}{f^4+f^2-E^2}}-1\right )~-f_0\right )\nonumber\\[2mm]
&&~-2q~\mbox{log}\Big(\frac{1-\mbox{cos}\gamma_j}{\mbox{sin}^2\gamma_j}\Big )~+~\mbox{{\tiny \cal O}}(1)~.\label{A1final}
\eea
\section*{Appendix D}
In this appendix we give arguments for the ansatz for the coordinates 
\eqref{u-rho-phi},\eqref{s-asympt} based on the experience with the explicit 
example in section 2. If we would take $s(\varphi)=0$, the variable $\varphi$
would have the meaning of an angle in the $(x_1,x_2)$-plane. Instead we want to argue, that in a well adapted coordinate system $\psi=\rho s(\varphi) +\varphi $ has this meaning. From the mapping pattern of section 2 we have \footnote{We keep the convention to denote the coordinates of the image by $y$'s.}
\beq 
\frac{y_1-1/q}{y_2}~=~\mbox{cot}(\pi- \gamma_1-\psi)~=~-\mbox{cot}(\gamma_1+\psi)~.
\eeq
On the other side the explicit mapping formulas \eqref{mapped-surface} yield
\beq
\frac{y_1-1/q}{y_2}~=~-~\mbox{cot}(\gamma_1+\varphi)~-~\frac{\rho(1+1/f^2)}{q~\sin(\gamma_1+\varphi)}~.
\eeq
Using \eqref{F-expansion} a comparison of these two expressions gives
\beq
\psi ~=~-\Big (\frac{3}{E}\Big )^{2/3}\frac{\rho}{q}~\big (\varphi ^{2/3}+{\cal O}(\varphi)\big )~+~\varphi ~+~{\cal O}(\rho, \varphi ^{4/3})~.
\eeq
Note that $\rho$ as used in section 2 is proportional to the radius variable in the sense of section 3 up to higher order corrections, which are not essential in this discussion, whose  only purpose is to see the emergence of the structure $\varphi + \mbox{ factor}\times \rho \varphi^{2/3}$.\\

\begin{figure}[h!]
 \centering
 \includegraphics[width=11cm]{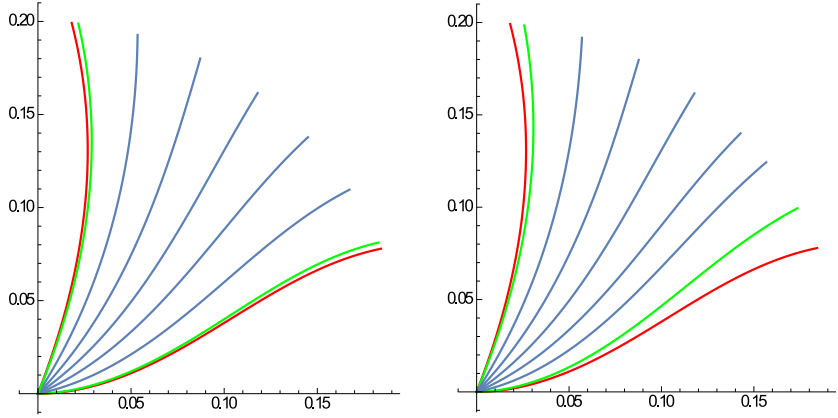} 
\caption{\it Example for $\theta=1.2$ and $\phi_1(\rho)=5\rho -15\rho ^2~,~~\phi_2=1.2+\rho+2\rho^2$ (red curves). Lines of constant $\varphi$ in step size 0.2 are shown in blue. Green lines show the first steps with size 0.02. On the left we see the situation for $s=0$, on the right for $s(\varphi):=a_1\varphi^{2/3}(1-\varphi/\theta)^{10}+a_2(\theta-\varphi)^{2/3}(\varphi/\theta)^{10}$. $a_1$,$a_2$ are adapted as described in the text. }
\label{new-coordinate-plot}
\end{figure}

To get an impression of the effect of working with $s(\varphi)$ as described in section 3 for a generic example, i.e. not related to section 2, we have added figure \ref{new-coordinate-plot}.\\[2mm] 

\section*{Appendix E}
Here we want to comment on some properties of the differential equation for $F_2(\varphi)$, see \eqref{nextleading-eom}.
As mentioned in the main text, for $F_1$ in \eqref{M}-\eqref{G1} we can take $F$ from appendix B. Furthermore, with \eqref{E-f},\eqref{f-F} we get
\beq
F'(\varphi )~=~\pm \frac{\sqrt{(1+F^2)(1+F^2-E^2F^4)}}{EF^2}\label{F'}~,~~~~F''(\varphi)~=~-\frac{2+2F^2+E^2F^6}{E^2F^5}~.
\eeq
The plus sign for $F'$ applies to $\varphi\in (0,\theta /2)$ and the minus to  $\varphi\in (\theta /2,\theta)$. 
The relation between $\varphi $ and $F$ is one to one in both half-intervals. The constant $E$ is related
to the maximum of $F$ (i.e. $F_0=F(\theta/2)=1/f_0$) via
\beq
E~=~\frac{\sqrt{1+F_0^2}}{F_0^2}~.
\eeq
Due to $F(\theta-\varphi)=F(\varphi)$ we have $G(\theta-\varphi)=G(\varphi)$ and $G_1(\theta-\varphi)=-G(\varphi)$. Thus the homogeneous equation related to \eqref{nextleading-eom} has both a symmetric as well as an antisymmetric
solution. The equation is regular  inside the whole interval $\varphi \in (0,\theta)$, but is singular
at its boundary points. 

Via
\beq
\tilde F_2(F)~:=~F_2(\varphi)\label{Ftilde}
\eeq
we can replace \eqref{nextleading-eom} by a differential equation with respect to $F$
\beq
\tilde F''_2(F)~+~g_1(F)\tilde F'_2(F)~+~g(F)\tilde F_2(F)~+~m(F)~=~0~,\label{nextleading-eom-mod}
\eeq
with \footnote{Note, that  $g$ and $g_1$ have the same form both for $\varphi\in (0,\theta/2)$ and $\varphi\in (\theta/2,\theta)$ since $F'$ enters quadratically in the calculation of $g$ and $g_1$ from $G$ and $G_1$. Of course this
symmetry between the two half-intervals for $\varphi$ is broken by $m(F)$ in the generic case.} 
\bea
g(F)&=&\frac{2+10F^2+6F^4-(2+10E^2)F^6}{(F+F^3)^2(E^2F^4-1-F^2)}~,\nonumber\\[2mm]
g_1(F)&=&\frac{2-F^4\big (2-E^2(F^2-4)\big )}{(F+F^3)(1+F^2-E^2F^4)}~,\nonumber \\[2mm]
m(F)&=&\frac{M\big (\varphi(F)\big )}{(dF/d\varphi )^2}~.
\eea
Remarkably, the coefficient functions of the homogeneous equation are now explicit known rational functions of $F$. Besides other benefits, this considerably simplifies the evaluation of numerical solutions. 

The equation \eqref{nextleading-eom-mod} is regular inside $F\in(0,F_0)$ and has regular singularities
at the boundaries. The one at $F_0$ is an artefact of the change of variables from $\varphi$ to $F$.
The corresponding Frobenius series for the homogeneous equation are power series in $\sqrt{F_0-F}$ or in $(F_0-F)$, related to odd or even power series in $(\varphi -\theta/2)$ for $F_2(\varphi)$.

Near $F=0$ we have $g(F)=-2 /F^2\times (1+{\cal O}(F^2))$ and $g_1(F)=2/F \times  (1+{\cal O}(F^2))$.
Therefore, the indicial equation for solving the homogeneous equation by the Frobenius method is $\alpha (\alpha -1)+2\alpha -2=0$.
Its solutions are $\alpha =-2$ and $\alpha =1$, corresponding to $\tilde F_2(F)=F^{-2}(1+\dots)$ and
$\tilde F_2(F)=F(1+\dots)$, where in both cases the dots stand for power series in $F^2$. 

What does this imply for the original differential equation w.r.t. $\varphi$, i.e. \eqref{nextleading-eom}? The just discussed behaviour at $F\rightarrow 0$ fixes via \eqref{F-expansion}-\eqref{f-F} the behaviour of the solutions of the homogeneous version of \eqref{nextleading-eom}, 
both at $\varphi\rightarrow 0$ and $\varphi\rightarrow\theta$. On both boundary points of the $\varphi$-interval we have a vanishing and a diverging solution. From the previous discussion of the behaviour
around the midpoint $\theta/2$ we also know, that there are symmetric as well as antisymmetric solutions.
However, a symmetric or antisymmetric solution, vanishing both at $\varphi=0$ and $\varphi =\theta$, could only exist, if
the spectrum of eigenvalues of the differential operator in \eqref{nextleading-eom} would contain the
value zero. Certainly, if at all, this could happen only at special values of the parameter
$F_0$, which contains the information on $\theta$. 

The upshot of this discussion so far is, that for a generic value of $\theta$ the homogeneous equation related to \eqref{nextleading-eom} has two independent solutions $y_1(\varphi)$ and $y_2(\varphi)$
with
\bea
y_1(\varphi)&\propto &\varphi ^{-\frac{2}{3}}~,~~~~~~~~~~y_2(\varphi)~\propto ~\varphi ^{\frac{1}{3}}~,~~~~~~~~~~~~~~~~\varphi\rightarrow 0\nonumber\\
y_1(\varphi)&\propto& (\theta-\varphi) ^{\frac{1}{3}}~,~~~~y_2(\varphi)~\propto ~(\theta-\varphi) ^{-\frac{2}{3}}~,~~~~~~~~\varphi\rightarrow \theta~.\label{y-asympt}
\eea
Let us now proceed to the construction of a solution for the full inhomogeneous equation  \eqref{nextleading-eom} by the method of varying the constants, i.e. starting with the ansatz
\beq
F_2(\varphi)~=~v_1(\varphi)~y_1(\varphi)~+~~v_2(\varphi)~y_2(\varphi)~.
\eeq
Then the coefficient functions $v_j(\varphi)$ have to obey the equations
\beq
v'_1(\varphi)~=~\frac{M(\varphi)}{W(\varphi)}~ y_2(\varphi)~,~~~~~~~~v'_2(\varphi)~=~-\frac{M(\varphi)}{W(\varphi)}~ y_1(\varphi)~.
\eeq
Their solutions are
\beq
v_1(\varphi )~=~v_1(0)~+~\int_0^{\varphi}\frac{M}{W}~y_2 d\varphi~,~~~v_2(\varphi )~=~v_2(0)~-~\int_0^{\varphi}\frac{M}{W}~y_2 d\varphi~.\label{v1v2}
\eeq
Here $W(\varphi)$  denotes the Wronskian. It is given by
\beq
W(\varphi)~=~\mbox{exp}\left (-\int ^{\varphi}G_1~d\varphi\right )~=~\mbox{const}\cdot\frac{(1+F^2)^3}{F^4}~.\label{wronski}
\eeq
From \eqref{wronski},\eqref{F-expansion} and \eqref{M-asympt} with \eqref{A1c1} and its analog for $\varphi\rightarrow\theta$ we see that the quotient $M/W$ at both boundary points of the $\varphi$-interval tends to nonzero constants. Therefore we get from \eqref{v1v2} and \eqref{y-asympt} near $\varphi =0$ and $\varphi =\theta$ 
\bea
F_2(\varphi)&=&v_1(0) (\varphi ^{-\frac{2}{3}}+\dots )+v_2(0) (\varphi ^{\frac{1}{3}}+\dots )+(b_1+b_2)(\varphi ^{\frac{2}{3}}+\dots)~,\\[2mm]
F_2(\varphi)&=&v_1(\theta) \big ((\theta-\varphi) ^{\frac{1}{3}}+\dots \big )+v_2(\theta) \big ((\theta-\varphi\big) ^{-\frac{2}{3}}+\dots )+(\hat b_1+\hat b_2)\big ((\theta-\varphi) ^{\frac{2}{3}}+\dots\big ).\nonumber
\eea
The $b_j$ and  $\hat b_j$ are some constants.

Now vanishing $F_2$ at both ends of the $\varphi$-interval requires
\beq
v_1(0)~=~v_2(\theta)~=~0~.
\eeq
Using \eqref{v1v2} this can be translated in conditions exclusively formulated at
one and the same endpoint, e.g. at $\varphi =0$
\beq
v_1(0)~=~0~,~~~~~v_2(0)~=~\int _0^{\theta}\frac{M}{W}~y_1~d\varphi ~.
\eeq
Note that $v_1(0)$ and  $v_2(0)$ are just the constants $B_1$ and $B_2$ in formula
\eqref{asF2} of the main text.
\section*{Appendix F}
Here we discuss the asymptotic evaluation of $A_{\epsilon,1}^{\tiny \mbox{c},0}(\rho _0)$ needed in formula \eqref{cusp-1} of the main text. One of our aim is the
identification of $\Gamma_{\mbox{\tiny cusp}}(\theta)$, defined in 
equation \eqref{gamma-cusp} of appendix B. Therefore, we replace the integration
variable $\varphi$ by $f=1/F$ as used in that appendix \footnote{Again $F_1$ of the main text can be identified with $F$ in appendix B.}
\beq
L_1(\varphi)~d\varphi ~=~\sqrt{\frac{f^4+f^2}{f^4+f^2-E^2}}~df~.
\eeq
Then 
\bea
~A_{\epsilon,1}^{\tiny \mbox{c},0}(\rho _0)&=&\int _{\rho <\rho_0,~~\rho/f+\rho ^2F_2+{\cal O}(\rho^3)>\epsilon}~ \frac{1}{\rho}~\sqrt{\frac{f^4+f^2}{f^4+f^2-E^2}~}~d\rho df\nonumber \\[3mm]
&=&\int_{f_0}^{f_{\mbox{\tiny max}}(\epsilon)}\left (\sqrt{\frac{f^4+f^2}{f^4+f^2-E^2}~}-1\right )df\int_{\rho_{\epsilon}(f)}^{\rho _0}\frac{d\rho}{\rho}\nonumber\\[3mm]
&&+\int_{f_0}^{f_{\mbox{\tiny max}}(\epsilon)}df\int_{\rho_{\epsilon}(f)}^{\rho _0}\frac{d\rho}{\rho}~, \label{I1I2}
\eea
with
\beq
\rho_{\epsilon}(f)~=~\frac{\sqrt{a^2+4\epsilon~ afB_2~}-a}{2B_2}~,~~~f_{\mbox{\tiny max}}(\epsilon)~=~\frac{\rho_0+\frac{B_2}{a}\rho_0^2+{\cal O}(\rho_0^3)}{\epsilon}~.
\eeq
Use has been made of the asymptotics of $F_2$ for small $\varphi$, i.e. large $f$, see \eqref{asF2},\eqref{F-expansion} and \eqref{asF1}.

Let us call the second and third line of \eqref{I1I2} $I_1$ and $I_2$ respectively. Then
\beq
I_1~=~\int_{f_0}^{f_{\mbox{\tiny max}}(\epsilon)}df~\left (\sqrt{\frac{f^4+f^2}{f^4+f^2-E^2}~}-1\right )~\mbox{log}\frac{2\rho_0B_2}{\sqrt{a^2+4\epsilon aB_2f~}-a}~.
\eeq
For $f$ fixed one has $\rho_{\epsilon}(f)=\epsilon f +{\cal O}(\epsilon ^2)$. However, since the upper integration boundary is $\propto 1/\epsilon$, one has to use 
this simplification with caution. In the case of $I_1$ the factor multiplying
the logarithm fastly goes to zero $\propto 1/f^4$, ensuring the finiteness of the integral and allowing the use of the simplified expression for  $\rho_{\epsilon}(f)$. This yields
\beq
I_1~=~\int_{f_0}^{\infty}df~\left (\sqrt{\frac{f^4+f^2}{f^4+f^2-E^2}~}-1\right )~\mbox{log}\frac{\rho_0}{\epsilon}~+~{\cal O}(1)~.
\eeq 

$I_2$ diverges for $\epsilon\rightarrow 0$. Here the use of the same simplification
for  $\rho_{\epsilon}(f)$ is not justified. But due to its simpler total integrand it can be performed explicitly
\bea
I_2&=&(f_{\mbox{\tiny max}}(\epsilon)-f_0)~\log \Big (\frac{2\rho_0B_2}{a}\Big)~-~\int_{f_0}^{f_{\mbox{\tiny max}}(\epsilon)}df~\log\Big (\sqrt{1+\frac{4\epsilon B_2}{a}~f~}-1 \Big )\nonumber\\
&&~=~\frac{1}{\epsilon}~\Big (\rho_0+\frac{B_2}{2a}\rho _0^2+{\cal O}(\rho_0^3)\Big )~+~f_0~\log\epsilon~+~{\cal O}(1)~.
\eea
The sum $I_1+I_2$ is then (with use of \eqref{gamma-cusp})
\beq
A_{\epsilon,1}^{\tiny \mbox{c},0}(\rho _0)~=~\frac{1}{\epsilon}~\Big( \rho_0+\frac{B_2}{2a}\rho _0^2+{\cal O}(\rho_0^3)\Big )~+~\frac{1}{2}~\Gamma _{\mbox{\tiny cusp}}(\theta)~\mbox{log}\epsilon ~+~{\cal O}(1)~.\label{Aeps1}
\eeq

 
\end{document}